\documentclass[english,sort&compress,superscriptaddress,aip,apl,reprint]{revtex4-1}
\usepackage{lmodern}
\usepackage[T1]{fontenc}
\usepackage[utf8]{inputenc}
\usepackage{amsbsy}
\usepackage{graphicx}

\makeatletter
\usepackage{hyperref}
\usepackage{amssymb,amsmath}

\makeatother

\usepackage{babel}
\begin{document}

\title{Equilibrium shape of single-layer hexagonal boron nitride islands
on low-index metal substrates}

\author{Marin Petrović}
\email{mpetrovic@ifs.hr}

\address{Faculty of Physics and CENIDE, University of Duisburg-Essen, Lotharstr. 1,
D-47057 Duisburg, Germany}

\affiliation{Center of Excellence for Advanced Materials and Sensing Devices,
Institute of Physics, Bijenička cesta 46, HR-10000 Zagreb, Croatia}

\author{Michael Horn-von Hoegen}

\affiliation{Faculty of Physics and CENIDE, University of Duisburg-Essen, Lotharstr. 1,
D-47057 Duisburg, Germany}

\author{Frank-J. Meyer zu Heringdorf}

\affiliation{Faculty of Physics and CENIDE, University of Duisburg-Essen, Lotharstr. 1,
D-47057 Duisburg, Germany}
\begin{abstract}
Large, high-quality layers of hexagonal boron nitride (hBN) are a
prerequisite for further advancement in scientific investigation and
technological utilization of this exceptional 2D material. Here we
address this demand by investigating chemical vapor deposition synthesis
of hBN on an Ir(111) substrate, and focus on the substrate morphology,
more specifically mono-atomic steps that are always present on all
catalytic surfaces of practical use. From low-energy electron microscopy
and atomic force microscopy data, we are able to set up an extended
Wulff construction scheme and provide a clear elaboration of different
interactions governing the equilibrium shapes of the growing hBN islands
that deviate from the idealistic triangular form. Most importantly,
intrinsic hBN edge energy and interaction with the iridium step edges
are examined separately, revealing in such way the importance of substrate
step morphology for the island structure and the overall quality of
2D materials.
\end{abstract}
\maketitle
Production of single- and multi-layer hexagonal boron nitride (hBN)
samples with minimum amount of defects has developed into one of the
most important areas of investigation of this insulating 2D material
(2DM) exhibiting high chemical stability and excellent thermal conductivity
\cite{Nagashima1995a,Pakdel2014,Zhang2017g}. Elimination of defects
from the production process is essential for scalable, high-throughput
synthesis of hBN that holds a great potential for advancements in
various fields of technology, such as field effect transistors \cite{Roy2014},
light-emitting diodes \cite{Ross2014} and sensors \cite{Sajjad2013}.
The method enabling such synthesis is chemical vapor deposition (CVD),
which in the case of hBN typically consists of initial nucleation
of individual islands on a catalyst metal surface, followed by island
growth and coalescence to form a full monolayer \cite{Kim2012c}.
When neighboring islands merge, defects are formed at the boundary,
resulting in lower material quality and deterioration of device performance
\cite{Gibb2013,Li2015f}. Since island coalescence is an unavoidable
step in CVD synthesis, it is important to understand all aspects of
island nucleation, shape, and growth, in order to develop new routes
for synthesis optimization.

Single-layer hBN has been grown via CVD on a wide range of single-
and poly-crystalline metal substrates, e.g., on Ru, Rh, Ni, Ir, Pd,
Pt, Cu and Fe \cite{Goriachko2007,Corso2004,Auwarter2003,Lee2012b,Orlando2011,Morscher2006,Cavar2008,Kim2012c,Joshi2012,Vinogradov2012b}.
Initially, hBN islands are often zig-zag (ZZ) terminated triangles,
with possible exceptions for some growth conditions \cite{Stehle2015,Poelsema2019,Felter2019}.
The triangles exhibit two dominant orientations, which originate from
the bi-elemental hBN unit cell \cite{Auwarter2003}. Further evolution
of island shape, and therefore the domain boundaries later on, can
be altered during CVD by adjusting the accessible parameters (e.g.,
temperature or precursor pressure/flux \cite{Stehle2015}), but the
choice of a particular substrate with its specific morphology is a
crucial initial factor, since precursor-substrate and hBN-substrate
interactions dictate the course of the synthesis.

A very important feature of substrate morphology are surface steps.
They are always present on both single-crystalline and poly-crystalline
foil substrates, and are often sites of hBN nucleation \cite{Sutter2008,Petrovic2017c,FarwickZumHagen2016}.
Also, island growth anisotropy induced by the substrate steps has
been observed for hBN on Ru(0001) \cite{Sutter2011a}, Pt(111) \cite{Zhang2015c},
Ir(111) \cite{Petrovic2017c}, and Cu(110) \cite{Wang2019e}, and
in a similar manner for graphene on metals \cite{Sutter2008,Loginova2009a,Wang2016}.
Due to the increased binding of the edge of 2DMs to the substrate
step edge, step-up (and in some cases also step-down) growth is hindered
and causes anisotropic growth rates of the islands. However, up to
now very little attention has been devoted to disentangling the different
energy factors that contribute to the observed island shapes. Here,
we elucidate the origin of hBN island shape anisotropy on metals by
performing a case study of hBN growth on Ir(111) with low-energy electron
microscopy (LEEM). We explicitly consider the intrinsic energy of
hBN island edges \cite{Liu2011d}, the islands' binding to the substrate,
and the specifics of the interaction between hBN island and substrate
step edges. Such detailed study is feasible because hBN-Ir interaction
is weak enough to allow hBN growth over the step edges, but is sufficiently
strong \cite{Preobrajenski2007} so that the effects of hBN edge-Ir
step edge interaction become clearly visible in experiments in the
form of distinct triangular and trapezoidal islands with ZZ edges
\cite{Petrovic2017c}.

Single-layer hBN was grown on Ir(111) in an ultra-high vacuum setup
via CVD by using borazine as a precursor. The Ir single-crystal was
cleaned by Ar sputtering at 2 keV followed by heating in oxygen at
1170 K and annealing at 1470 K. Unless otherwise noted, the borazine
pressure during CVD was $10^{-8}$ mbar and the temperature was 1170
K. An Elmitec SPE-LEEM III microscope was used to carry out \textit{in-situ},
bright field LEEM and selected-area low-energy electron ($\mu$-LEED)
measurements. Atomic force microscope (AFM) measurements were performed
\textit{ex-situ} in air with a Veeco Dimension 3100 microscope operated
in tapping mode.

\begin{figure}[!h]
\begin{centering}
\includegraphics{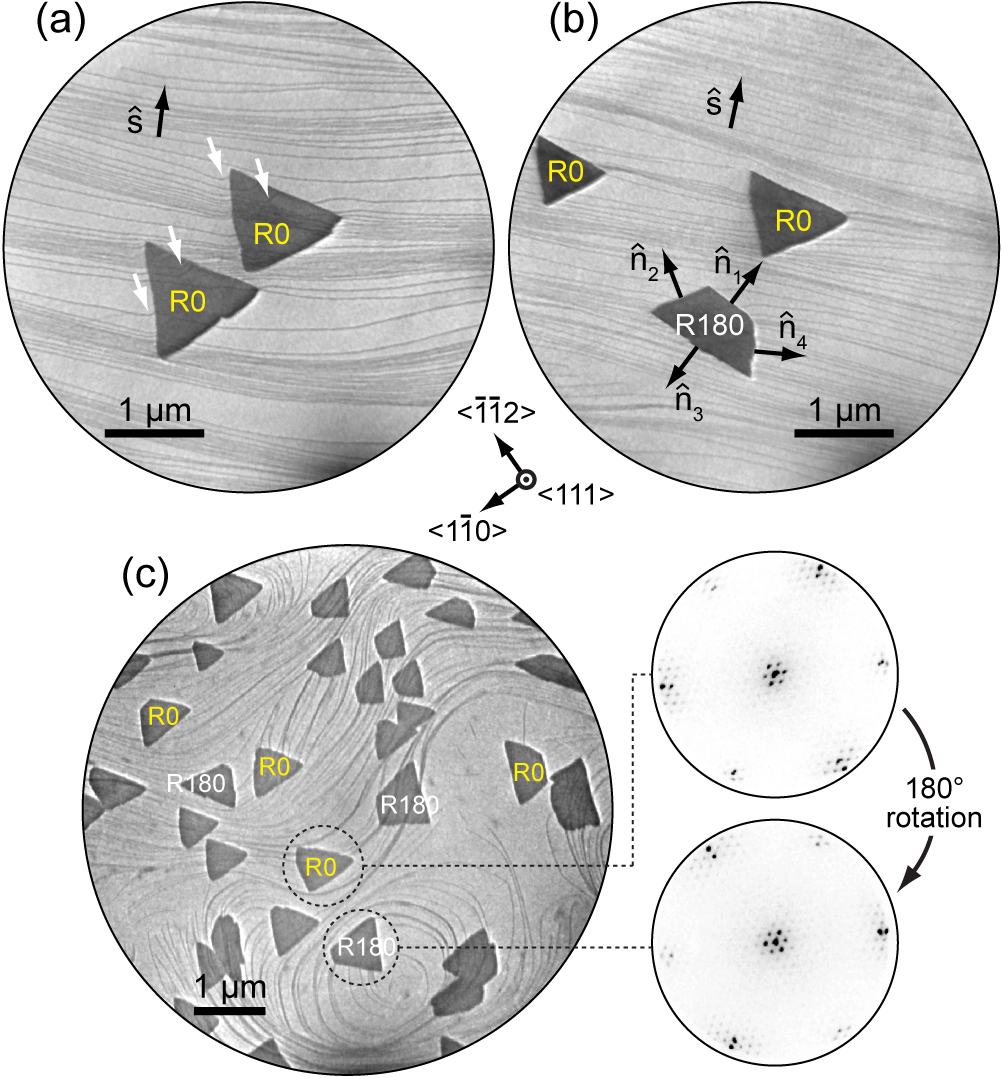}
\par\end{centering}
\caption{\label{fig1}(a) and (b) LEEM images of hBN islands on Ir(111) with
uniform $\boldsymbol{\hat{s}}$ (designating Ir step-up direction).
White arrows indicate sites of strong Ir step edge bending. Edge normals
\textbf{$\boldsymbol{\hat{n}}_{\mathrm{i}}$} are noted for one of
the islands in panel (b). (c) LEEM image of R0 and R180 islands, including
two representative $\mu$-LEED patterns, on a part of the Ir surface
with strongly varying $\boldsymbol{\hat{s}}$. (a) and (b) $E=24.1$
eV, (c) $E=17.5$ eV in LEEM and $E=35.2$ eV in $\mu$-LEED. Crystallographic
directions noted in the center of the figure apply to all LEEM and
LEED images.}
\end{figure}

LEEM images in Figs. \ref{fig1}(a) and (b) show isolated hBN islands
on the Ir surface, where thin dark lines spanning across the field
of view are Ir step edges. Crystallographic analysis, taking into
account image rotation in LEEM (e.g., see Ref. \onlinecite{Felter2019}
for technical details), reveals that the hBN edges are of ZZ type.
The orientation of hBN island edges can be described by unit vectors
\textbf{$\boldsymbol{\hat{n}}$} which are perpendicular to the edges,
as shown in Fig. \ref{fig1}(b). The local orientation of Ir steps
is designated by a unit vector $\boldsymbol{\hat{s}}$ which for every
point along the step is perpendicular to the step. The orientation
of \textbf{$\boldsymbol{\hat{s}}$}, i.e., the difference between
step-up and step-down direction, can be determined by recognizing
that the short base of the trapezoid must be facing the Ir step-up
direction (see AFM data below). When straight steps are present on
the Ir surface and \textbf{$\boldsymbol{\hat{s}}$} shows minor change
across a large area, as in Fig. \ref{fig1}(a) and (b), one hBN orientation
(denoted as R0) grows exclusively in triangular form, and the other
one (rotated by $180^{\circ}$, denoted as R180) grows exclusively
in the form of trapezoids, as we reported earlier \cite{Petrovic2017c}.
The shape of islands changes, however, when the Ir substrate exhibits
a complex surface morphology and contains step edges with large curvature,
including hills and valleys. Such a situation is shown in Fig. \ref{fig1}(c),
where many R0 and R180 islands are visible (in this particular case,
the borazine pressure during synthesis was $6\times10^{-8}$ mbar,
leading to a higher island density). Whether the islands are of R0
or R180 orientation can be deduced (i) by simply measuring the orientation
of their edges, or (ii) by comparing the islands' 3-fold symmetric
$\mu$-LEED patterns as illustrated in Fig. \ref{fig1}(c).

A careful inspection of LEEM images reveals that R0 islands are triangular
on some parts of the surface, while they are trapezoidal on other
parts {[}see labeled islands in Fig. \ref{fig1}(c){]}. Moreover,
the short base of trapezoidal islands is found at different positions,
i.e., the triangles truncation occurs at different vertices in order
to form trapezoids. This is also valid for R180 islands. Considering
that the short base of trapezoidal islands faces the step-up direction
of Ir, we deduce that R0 and R180 islands are not predetermined to
grow as triangles or trapezoids, but their shape is governed by Ir
step morphology.

\begin{figure}
\begin{centering}
\includegraphics{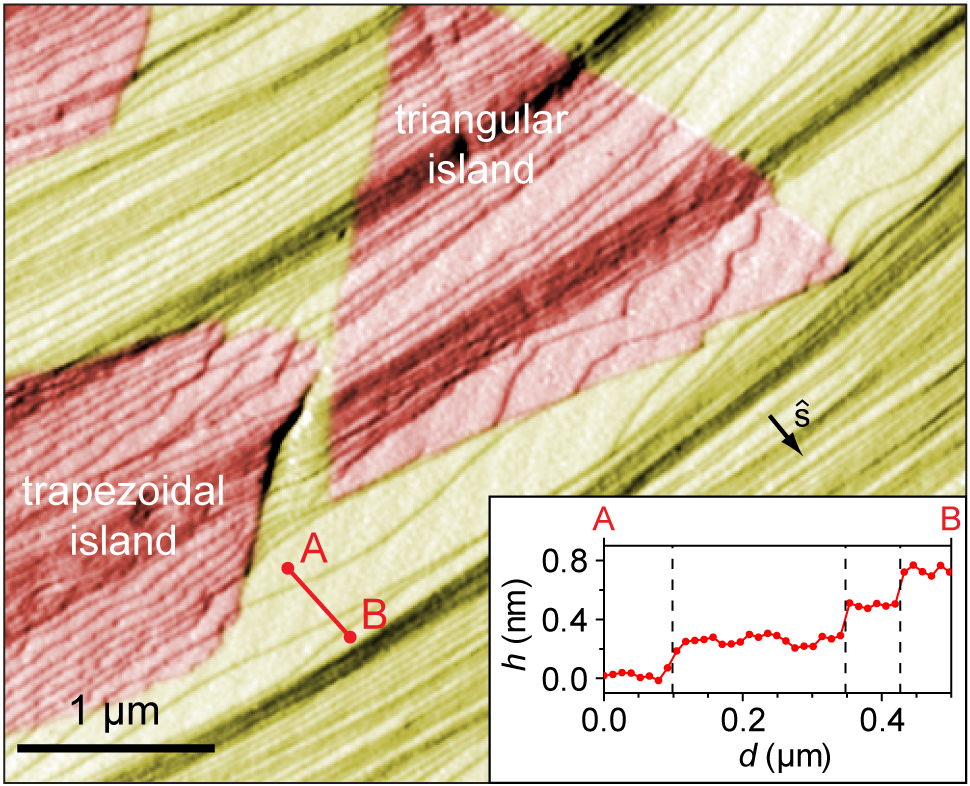}
\par\end{centering}
\caption{\label{fig2}AFM image (first derivative of topography in $x$) of
hBN islands (red) on Ir(111) (yellow). Thin diagonal lines are Ir
step edges. The inset shows line profile between points $A$ and $B$
prior to AFM image differentiation, vertical dashed lines mark positions
of Ir step edges.}
\end{figure}

\begin{figure*}
\begin{centering}
\includegraphics{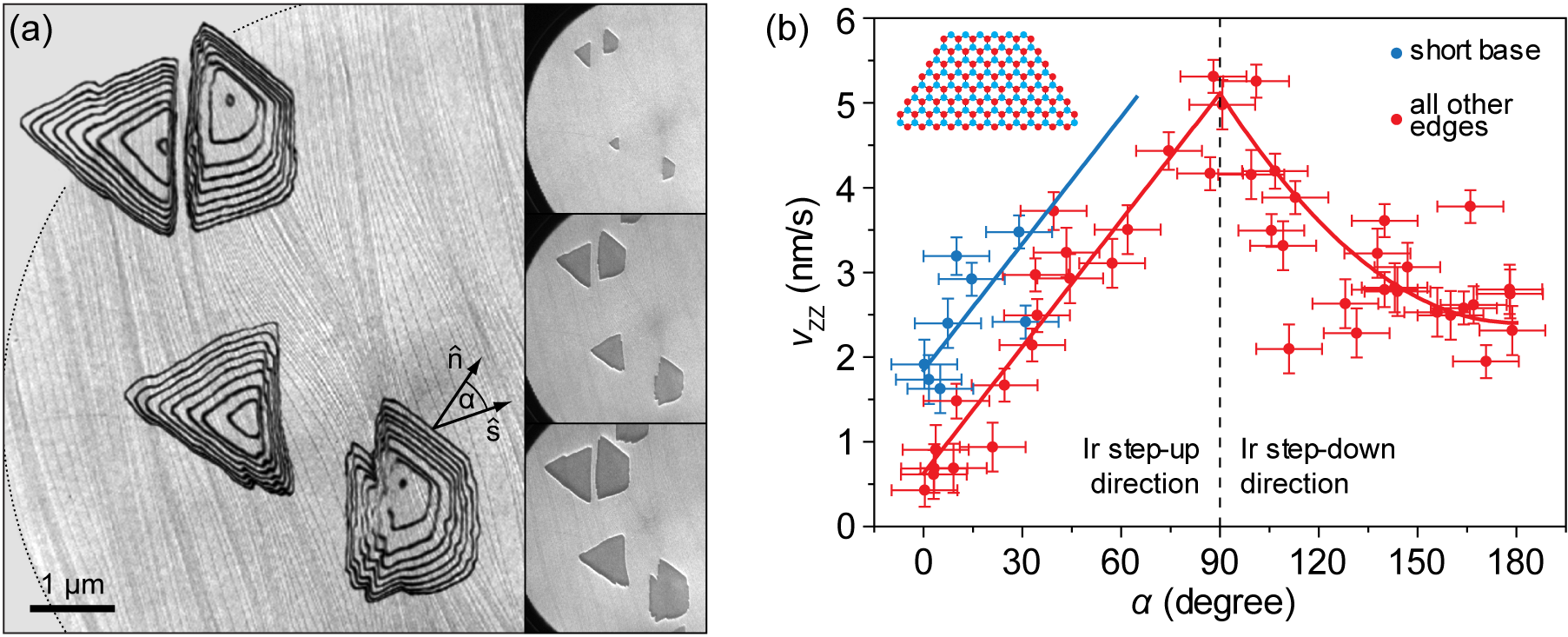}
\par\end{centering}
\caption{\label{fig3}(a) Stacked contours of hBN islands extracted from a
LEEM growth sequence, three characteristic images are shown in the
panels on the right. Contours correspond to islands perimeters recorded
at time intervals of 32 s. Edge normal $\boldsymbol{\hat{n}}$, local
step-up direction $\boldsymbol{\hat{s}}$ and an angle between them
$\alpha$ are noted for one hBN edge. $E=17.3$ eV. (b) Data plot
of ZZ edge advancement speed $v_{\mathrm{ZZ}}$ as a function of angle
$\alpha$. Red and blue lines are the fits to the data.}
\end{figure*}

Close to the hBN edges, Ir steps are often strongly bent, as marked
by white arrows in Fig. \ref{fig1}(a), indicating that the interaction
between hBN island edges and Ir step edges is significant and plays
an important role in the growth of hBN. Recently, Poelsema et al.
also found that the growth of hBN on Ir(111) results in a severe reorientation
of Ir step edges \cite{Poelsema2019b}. A better view of step layout
can be obtained from scanning probe imaging. In Fig. \ref{fig2},
an AFM image with several hBN islands is shown. The step-up direction
is easily identified from an AFM profile shown in the inset. Ir step
edges, which are rather straight in the hBN-free region, are distorted
in areas where hBN islands overgrew them. This is most prominent at
the lower right edge of the triangular island that faces the Ir step-up
direction, where Ir steps exhibit a wavy structure and contain straight
segments parallel to the hBN edge. Furthermore, in Fig. \ref{fig2},
a short base of the trapezoidal island has formed in the step-up direction,
facilitating in such way parallel configuration of hBN edge and Ir
steps. Our AFM data suggests that strong hBN-Ir interaction favors
attachment of hBN edges to Ir step edges which is enabled by their
parallel alignment, most prominently in the step-up direction of Ir.

We now proceed to discuss the evolution of the island shape during
growth. Contours of several hBN islands extracted from LEEM data are
shown in Fig. \ref{fig3}(a), corresponding to the ZZ-type perimeters
of R0 and R180 islands recorded at time intervals of 32 s. The island
growth is quantified by measuring edge distances from the island nucleation
site, $d$, and calculating the average advancement speed as $v_{\mathrm{ZZ}}=d/t$.
At the same time, the angle $\alpha$ between $\boldsymbol{\hat{n}}$
and \textbf{$\boldsymbol{\hat{s}}$} has been measured for each edge,
thus providing information to plot $v_{\mathrm{ZZ}}$ as a function
of $\alpha$. In total, 16 islands have been analyzed, and from the
data shown in Fig. \ref{fig3}(b) it is clear that hBN edges propagate
faster (slower) when their normals $\boldsymbol{\hat{n}}$ are perpendicular
(parallel) to the local direction of $\boldsymbol{\hat{s}}$. Short
bases of trapezoids (blue dots) have been singled out from all other
edges (red dots) because of their different elemental composition
(see inset). The data in Fig. \ref{fig3}(b) shows functional dependence,
and we fit it with simple mathematical functions in order to establish
an analytical model of hBN growth. At $\alpha=90^{\circ}$, the island
edge changes its growth orientation from step-up to step-down with
respect to the Ir surface, and it is reasonable to assume modification
of hBN-Ir interaction and also a different behavior of $v_{\mathrm{ZZ}}\left(\alpha\right)$.
Therefore, we fit the data with a combination of linear (for $\alpha\leq90^{\circ}$)
and quadratic (for $\alpha>90^{\circ}$) functions (see Supporting
Section S1 for fit details).

From a comparison of the product $v_{\mathrm{ZZ}}L$, where $L$ is
the typical island size, with the values of the diffusion coefficient
$D$ of various borazine fragments, i.e., the building blocks for
hBN growth, it is clear that $D\gg v_{\mathrm{ZZ}}L$ (see Supporting
Section S2 for more details). This implies that hBN growth on Ir(111)
is taking place near the thermodynamic equilibrium and that $v_{\mathrm{ZZ}}$
is proportional to the edge free energy \cite{Artyukhov2012}. In
that case, the thermodynamic Wulff construction can be applied to
obtain the shape of 2D islands \cite{Artyukhov2012,Zhang2016a}, and
hence we use it to reconstruct the observed R0 and R180 island forms.
We use an analytic expression for the epitaxial hBN island edge energy
per unit length as a function of polar angle and chemical potential
difference, $\gamma\left(\chi,\Delta\mu\right)$ \cite{Zhang2016a,Liu2011d}
(see Supporting Section S3 for details). The chemical potential $\Delta\mu$
is defined as a disbalance between chemical potentials of B and N
atoms, $\Delta\mu=\left(\mu_{\mathrm{B}}-\mu_{\mathrm{N}}\right)/2$.
Highly positive $\Delta\mu$ favors B-terminated ZZ edges, while highly
negative $\Delta\mu$ favors N-terminated ZZ edges. $\gamma\left(\chi,\Delta\mu\right)$
also contains binding energies of ZZ and armchair edges to Ir that
can be found in the literature \cite{Laskowski2008,Zhao2015,FarwickZumHagen2016}.

\begin{figure*}
\begin{centering}
\includegraphics{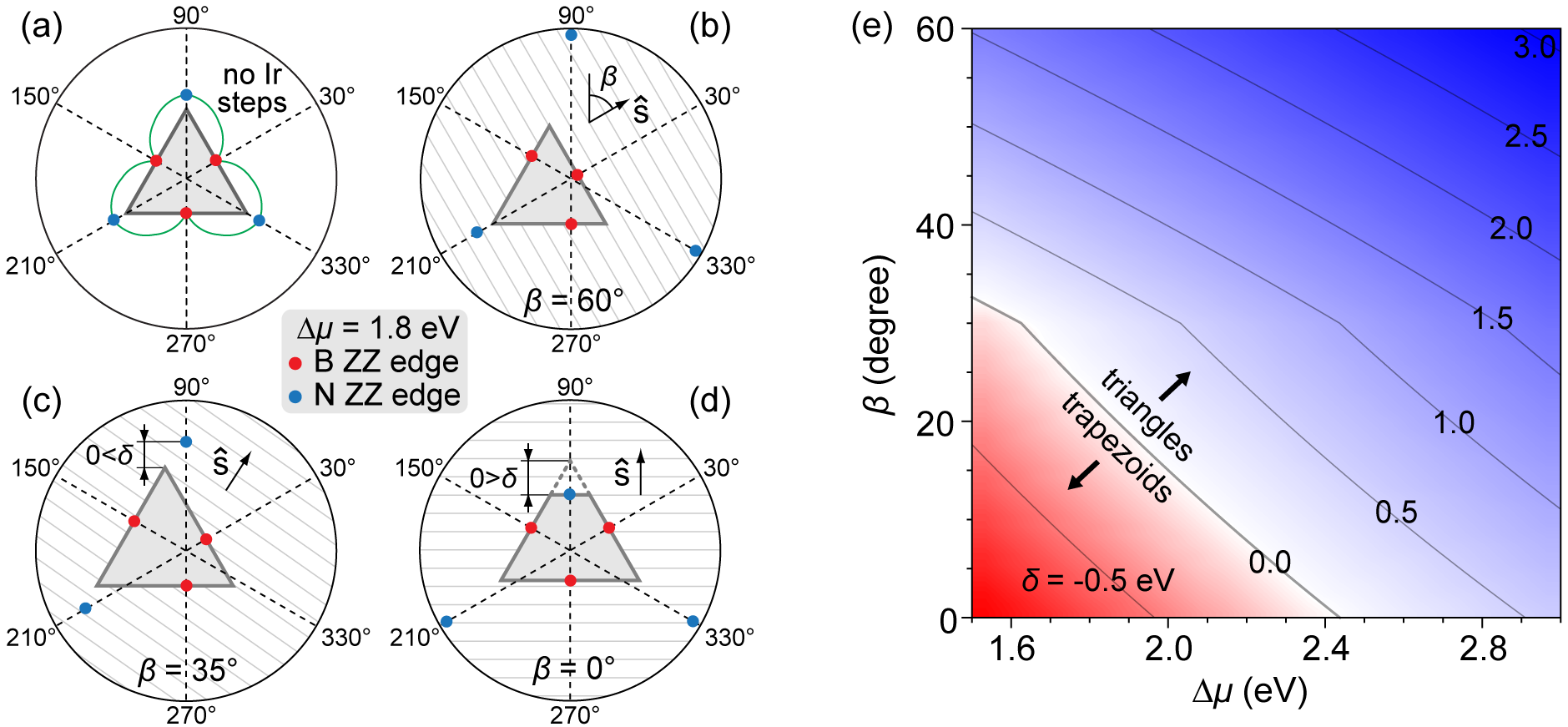}
\par\end{centering}
\caption{\label{fig4}(a) Thermodynamic Wulff construction of an hBN island
on Ir without steps. Green line is $\gamma\left(\chi,\Delta\mu\right)$,
and gray triangle indicates the Wulff shape. (b)-(d) Transition of
a triangular island shape to trapezoidal after including the interaction
of hBN with the Ir steps and subsequent rotation of $\boldsymbol{\hat{s}}$.
Gray lines indicate Ir steps. (e) Plot of $\delta\left(\beta,\Delta\mu\right)$.
For $\delta<0$ (red region) and $\delta>0$ (blue region), trapezoidal
and triangular islands grow on the Ir surface, respectively. }
\end{figure*}

To be consistent with the realistic experimental values of $\Delta\mu$
\cite{Zhao2015}, and also with the preference of B-terminated ZZ
edges \cite{FarwickZumHagen2016,Petrovic2017c}, our analysis is restricted
to 1.5 eV $<\Delta\mu<$ 3 eV. A comparable analysis with N-terminated
island edges, where $\Delta\mu<0$, is straightforward. The Wulff
construction of an hBN island at $\Delta\mu=1.8$ eV is shown in the
polar plot in Fig. \ref{fig4}(a). The edge energy $\gamma\left(\chi,\Delta\mu\right)$
used in this construction includes the hBN island's intrinsic edge
energy and the binding of hBN to the flat Ir substrate without steps
(details are given in the Supporting Section S3). The island shape
is determined by the red points of $\gamma\left(\chi,\Delta\mu\right)$
which correspond to B-terminated ZZ edges. N-terminated ZZ edges,
designated by blue points, have higher energy and therefore do not
constitute the edges of hBN island at these conditions. It follows
from Fig. \ref{fig3} and the outlined diffusion considerations that
$v_{\mathrm{ZZ}}\left(\alpha\right)$ corresponds to the ZZ edge energy
modulation arising from the relative edge orientation with respect
to the Ir steps. This is incorporated into the Wulff construction
by applying $\gamma\left(\chi,\Delta\mu\right)\rightarrow\gamma\left(\chi,\Delta\mu\right)\cdot v_{\mathrm{ZZ}}\left(\chi\right)$
and by repositioning the red and the blue points in Figs. \ref{fig4}(b)-(d)
accordingly. Our LEEM data shows that these few points are the only
relevant ones to describe the shape of hBN islands.

For the sake of clarity, we focus on the top vertex and the upper-right
edge of the triangular hBN island in Figs. \ref{fig4}(a)-(d) to examine
the truncation effect of the island. Introduction of Ir steps modifies
the energies of all hBN edges, depending on the orientation of hBN
island with respect to Ir steps. This orientation is quantified by
the angle $\beta$ measured between $\boldsymbol{\hat{s}}$ and direction
corresponding to $\chi=90^{\circ}$ {[}see Fig. \ref{fig4}(b){]}.
For relatively large values of $\beta$ as in Figs. \ref{fig4}(b)
and (c), Ir step-bending as visible in the AFM data of Fig. \ref{fig2}
is the optimal mechanism for hBN island energy minimization. The required
bending at the upper-right island edge is not large (step energy increases
with the step curvature \cite{Giesen2001}) and its cost is compensated
by an overall energy gain achieved by strong binding between parallel
hBN edge and Ir steps. In such a situation, B-terminated ZZ edges
remain energetically preferred. As $\beta$ decreases, the cost of
Ir step bending becomes too high and it becomes energetically non-profitable.
However, the energy of N-terminated ZZ edge at the top of the island
in Fig. \ref{fig4}(d) is reduced significantly since it becomes (nearly)
parallel with Ir steps, resulting in hBN island truncation at the
vertex pointing in the step-up direction. After truncation, at the
short base of the trapezoidal island, it takes much less (if any)
Ir step bending to achieve a parallel configuration and strong binding
between hBN island edge and Ir steps, and in such way an initially
unfavorable N-termination of hBN islands is energetically compensated.

The presented Wulff construction predicts, in agreement with our LEEM
data, that the truncation is allowed only in the Ir step-up direction.
It can be argued that in the step-up direction hBN edges become passivated
by binding to metal atoms. The result of such an interaction would
be a much stronger binding in the step-up direction as compared to
the step-down direction, similarly to the case of epitaxial graphene
\cite{Wang2016}. Binding of different hBN edges to Ir steps that
undergo bending and repositioning constitutes the energetic background
of $v_{\mathrm{ZZ}}\left(\alpha\right)$, and that is why the inclusion
of $v_{\mathrm{ZZ}}\left(\alpha\right)$ into the Wulff construction
is crucial for obtaining the experimentally observed shapes of hBN
islands.

A systematic investigation of the truncation effect is shown in Fig.
\ref{fig4}(e) in which $\delta$, i.e., the vertical separation between
N-terminated ZZ edge and its closest vertex as depicted in Figs. \ref{fig4}(c)
and (d), is plotted as a function of $\Delta\mu$ and $\beta$. For
certain $\left(\beta,\Delta\mu\right)$ combinations, trapezoidal
islands ($\delta<0,$ red region) are energetically preferred over
triangular ones ($\delta>0$, blue region). This explains why R0 and
R180 islands have different shapes on the surface with uniform $\boldsymbol{\hat{s}}$,
and also why do they change their shape when $\boldsymbol{\hat{s}}$
(i.e., $\beta$) changes. The level of truncation of the triangle
also depends on $\Delta\mu$ and \textbf{$\beta$}, explaining trapezoids
of different heights in Fig. \ref{fig1}(c).

In summary, we have shown that the step morphology of the substrate
used in CVD growth of hBN is a crucial factor which determines the
energetically most stable shape of synthesized hBN islands. The total
energy of the system is minimized by adhering hBN edges to the Ir
step edges, which is achieved by repositioning of Ir steps and formation
of trapezoids (instead of triangles) during hBN growth. The degree
of Ir step repositioning and the feasibility of trapezoidal shape
depend on the relative orientation between hBN island edges and Ir
steps, and the chemical potentials of B and N atoms during the synthesis
of hBN. The use of an extended Wulff construction allowed exact pinpointing
of different energy contributions governing hBN growth, and this enables
the application of our results, by adjusting the relevant interaction
parameters, in studies of hBN synthesis on other metal substrates.

See Supporting information for details of the $v_{\mathrm{ZZ}}\left(\alpha\right)$
data fit, evaluation of the diffusion coefficients, and details of
$\gamma\left(\chi,\Delta\mu\right)$.

The Alexander von Humboldt Foundation is acknowledged for financial
support. M.P. would like to thank Thomas Michely for stimulating discussions.

\bibliographystyle{apsrev4-1}
\bibliography{Collection4}

\end{document}